\documentclass{article}
\usepackage[utf8]{inputenc}
\usepackage{enumitem}
\usepackage{amsmath}
\usepackage{amssymb}
\usepackage{amsthm}
\usepackage{braket}
\usepackage[a4paper, total={6in, 9in}]{geometry}
\usepackage{xcolor}
\usepackage{hyperref}
\usepackage{titlesec}
\usepackage{comment}
\usepackage{soul}

\usepackage{authblk}

\usepackage[linesnumbered]{algorithm2e}
\usepackage{tikz}
\usetikzlibrary{decorations.pathreplacing}
\usetikzlibrary{calc}
\usetikzlibrary{quantikz2}
\usetikzlibrary[graphs, graphs.standard]
\usetikzlibrary{positioning,chains,fit,shapes,calc,decorations.markings}
\usetikzlibrary{decorations.pathreplacing,calligraphy}

\tikzstyle{vertex}=[circle, draw, inner sep=0pt, minimum size=6pt]
\tikzstyle{marked}=[circle,draw,dotted,inner sep=0pt, minimum size=6pt]

\newcommand{\abs}[1]{\left|#1\right|}
\newcommand{\matrx}[1]{\begin{bmatrix} #1 \end{bmatrix}}
\newcommand{\U}{\mathcal{U}}
\newcommand{\pr}[1]{\left( #1 \right)}
\newcommand{\im}{\textnormal{i}}
\newcommand{\paren}[1]{\left( #1 \right)}

\newcommand{\ketbra}[2]{|#1\rangle\langle#2|}
\renewcommand{\ket}[1]{\left| #1 \right\rangle}

\newcommand{\floor}[1]{{\left\lfloor #1 \right\rfloor}}
\newcommand{\ceil}[1]{{\left\lceil #1 \right\rceil}}

\renewcommand{\bar}[1]{\overline{#1}}

\titleclass{\subsubsubsection}{straight}[\subsection]

\newcounter{subsubsubsection}[subsubsection]
\renewcommand\thesubsubsubsection{\thesubsubsection.\arabic{subsubsubsection}}

\titleformat{\subsubsubsection}
  {\normalfont\normalsize\bfseries}{\thesubsubsubsection}{1em}{}
\titlespacing*{\subsubsubsection}
{0pt}{3.25ex plus 1ex minus .2ex}{1.5ex plus .2ex}

\makeatletter
\renewcommand\paragraph{\@startsection{paragraph}{5}{\z@}%
  {3.25ex \@plus1ex \@minus.2ex}%
  {-1em}%
  {\normalfont\normalsize\bfseries}}
\renewcommand\subparagraph{\@startsection{subparagraph}{6}{\parindent}%
  {3.25ex \@plus1ex \@minus .2ex}%
  {-1em}%
  {\normalfont\normalsize\bfseries}}
\def\toclevel@subsubsubsection{4}
\def\toclevel@paragraph{5}
\def\toclevel@paragraph{6}
\def\l@subsubsubsection{\@dottedtocline{4}{7em}{4em}}
\def\l@paragraph{\@dottedtocline{5}{10em}{5em}}
\def\l@subparagraph{\@dottedtocline{6}{14em}{6em}}
\makeatother

\newtheorem{theorem}{Theorem}
\newtheorem{corollary}{Corollary}[theorem]

\RestyleAlgo{ruled}

\SetKwInOut{KwIn}{Input}
\SetKwInOut{KwOut}{Output}

\title{Quantum Counting on the Complete Bipartite Graph}
\author[1]{Gustavo A. Bezerra}
\author[2]{Raqueline A. M. Santos}
\author[1]{Renato Portugal}
\affil[1]{\small National Laboratory of Scientific Computing, Petropolis, RJ, 25651-075, Brazil}
\affil[2]{\small Center for Quantum Computer Science, Faculty of Computing, University of Latvia, Raina bulvaris 19, LV-1586, Riga, Latvia}

\begin{document}
\maketitle

\begin{abstract}
Quantum counting is a key quantum algorithm that aims to determine the number of marked elements in a database. This algorithm is based on the quantum phase estimation algorithm and uses the evolution operator of Grover's algorithm because its non-trivial eigenvalues are dependent on the number of marked elements. Since Grover's algorithm can be viewed as a quantum walk on a complete graph, a natural way to extend quantum counting is to use the evolution operator of quantum-walk-based search on non-complete graphs instead of Grover's operator. In this paper, we explore this extension by analyzing the coined quantum walk on the complete bipartite graph with an arbitrary number of marked vertices. We show that some eigenvalues of the evolution operator depend on the number of marked vertices and using this fact we show that the quantum phase estimation can be used to obtain the number of marked vertices. The time complexity for estimating the number of marked vertices in the bipartite graph with our algorithm aligns closely with that of the original quantum counting algorithm.
\end{abstract}

\section{Introduction}

Quantum computers have the potential to perform certain tasks significantly faster than classical computers due to the principles of quantum mechanics~\cite{NC00}. This is demonstrated by quantum-walk-based search~\cite{portugal2018quantum} and quantum phase estimation algorithms~\cite{KSV02}. Quantum walk-based search algorithms are faster and more efficient than their classical counterparts for searching marked vertices on many graphs. On the other hand, the quantum phase estimation algorithm is a fundamental tool in quantum computing that enables the determination of the eigenvalues of a unitary operator, which is useful for various applications. These algorithms are just a few examples of the potential of quantum computing to provide faster and more efficient solutions to computational problems. They are important in our context, as we can merge them to solve the following problem: How can we find the number of marked vertices in a graph?

Boyer at al.~\cite{boyer1998tight} described a generalization of Grover's algorithm~\cite{grover1997quantum} with $k$ marked elements, which is able to find a marked element in $O\big(\sqrt{N/k}\big)$ steps if we know $k$ beforehand. The counting problem naturally arises from this generalization. The problem now is how we determine $k$, if this information is not provided beforehand. This problem was addressed in~\cite{brassard2002quantum,kaye2007introduction}, where the phase estimation algorithm is used (as a subroutine) with the goal of finding a non-trivial eigenvalue of the evolution operator of Grover's algorithm. This eigenvalue provides an estimate of $k$, that can be used to determine the number of iterations required by the search algorithm. 

Quantum walks, the quantum analog of classical random walks, play an important role in quantum computing and are a crucial component in the design of quantum search algorithms~\cite{SKW03,Amb07a}. Grover's algorithm can be described as a discrete-time quantum walk on the complete graph~\cite{AKR05,portugal2018quantum}. In early quantum-walk-based search algorithms, the coin operator was modified to invert the phase of the state when the walker is at a marked vertex. This modification essentially involves using the standard evolution operator $SC$,  where $S$ is the shift operator and $C$ is the coin operator, preceded by the application of an oracle $R$. The oracle $R$ reveals the information about which vertices are marked, thereby defining the evolution operator of the quantum-walk-based search as $U = S C R$.

In this work, we address the problem of counting the number of marked vertices within the framework of quantum-walk-based search algorithms. Here, we consider a discrete-time quantum walk on a graph, with $k$ marked vertices, and an oracle $R$ that inverts the phase of these marked vertices while leaving the unmarked ones unaffected. The counting problem aims to determine $k$ when it was not given beforehand. Unlike Grover's algorithm with multiple marked elements, the complexity of this problem increases significantly when applied to a graph with less symmetry than a complete graph. This is primarily due to the increased difficulty in determining the eigenvalues of the evolution operator.

Recently, Le Gall and Ng~\cite{gall2022quantum} used the same strategy to estimate the number of marked states for any reversible ergodic Markov chain. Our results specifically focuses on the complete bipartite graph, which corresponds to a non-ergodic Markov chain. Let the total number of marked vertices of the complete graph be $k=k_0+k_1$. Here, $k_0$ and $k_1$ represent the counts of marked vertices in the first and second parts of the graph, respectively. Our algorithm addresses this problem by initially counting the marked vertices in the first part, followed by those in the second. This is achieved by modifying the oracle using two controlled $Z$ gates acting on an auxiliary qubit. The time complexity of our approach is comparable to that of the original quantum counting algorithm as described in Ref.~\cite{brassard2002quantum}, which is $\Theta(\sqrt{N})$ for an allowed error of $\sqrt{N}$, where $N$ is the total number of vertices in the graph.
This work is an extension of Bezerra's Master's thesis~\cite{bezerra2021master}\footnote{Available at
    \url{https://tede.lncc.br/handle/tede/341?locale=en} and
    \url{https://arxiv.org/abs/2312.03768}.}.

This work is organized as follows: Section~\ref{sec1} provides a review of both the original quantum counting algorithm and Kitaev's phase estimation algorithm. In Section~\ref{sec:evol-op}, we discuss the evolution operator of a coined quantum walk search on the complete bipartite graph, along with its eigenvalues and eigenvectors. Section~\ref{sec:counting-complete-bipartite} details our algorithm for counting the number of marked vertices in a complete bipartite graph. Finally, Section~\ref{conclusion} offers our concluding remarks.

\section{Review on quantum counting}\label{sec1}

The phase estimation algorithm~\cite{kitaev1995quantum} is at the core of the
quantum counting algorithm developed by Brassard~{\it et al.}~\cite{brassard2002quantum}. In this section, we provide an overview of the phase estimation algorithm (section~\ref{sec:phase-est}), and we examine its application within the quantum counting algorithm (section~\ref{sec:orig-count}).

\subsection{Phase estimation algorithm}
\label{sec:phase-est}

The phase estimation algorithm is a fundamental quantum algorithm that determines the phase associated with a specific eigenvector of a unitary operator. It has significant relevance in many quantum computing applications, including order finding and factoring. 

The algorithm starts by preparing a quantum state which is an eigenvector of a unitary operator. Then it applies a series of controlled unitary operations on a set of ancilla qubits in a superposition state. This procedure leads to a phase kickback, encoding the phase information into the ancilla qubits. The quantum Fourier transform is applied to the ancilla qubits, which translates the phase information into a readable format. At the end, the measurement of these qubits provides an estimate of the phase. The precision of the estimation increases with the number of ancilla qubits used.

To be more precise, let $U$ be a unitary matrix and let $\ket{\theta_j}$ be an eigenvector of $U$ associated with eigenvalue $e^{\im\theta_j}$, where $0 \leq \theta_j < 2\pi$ is unknown. We refer to $\theta_j$ as an eigenphase of the unitary operator $U$. The phase estimation algorithm employs two registers.
When the input of the second register is $\ket{\theta_j}$,
a measurement in the first register yields a $p$-bit estimate of $P\theta_j/2\pi$,
and the algorithm outputs an estimate of $\theta_j$, denoted as $0 \leq \tilde{\theta_j} < 2\pi$.
Here, $p$ is the number of qubits in the first register, and $P = 2^p$. Typically, the input $\ket{\psi}$ of the second register is not an eigenvector of $U$. In this scenario, we express $\ket\psi = \sum_j \alpha_j \ket{\theta_j}$ in the eigenbasis of $U$, and the probability of estimating $\theta_j$ is $|\alpha_j|^2$. The phase estimation algorithm is outlined in Algorithm~\ref{alg:phase-estimation}.

\begin{algorithm}[!htb]
    \caption{\texttt{phase\_estimation($p, U, \ket\psi$)}}
    \label{alg:phase-estimation}   
    \KwIn{$\ket{0}^{\otimes p}\ket\psi$, where $p$ is the number of qubits of the first register and $\ket\psi$ belongs to the eigenspace of $U$.}
    \KwOut{$\tilde{\theta}$. }
    Prepare $\ket{\psi_0} = \ket{0}^{\otimes p}\ket\psi$ as the initial state\;
    Apply the Hadamard gate $H$ to each qubit in the first register\;
    Apply $U^{2^{p-j}}$ on the second register controlled by qubit $j$, for $1\le j\le p$\;
    Apply the inverse quantum Fourier transform to the first register\;
    Measure the first register
    in the computational basis (assume the result is $P \tilde\theta / 2\pi$)\;
    \Return $\tilde{\theta}$.
\end{algorithm}

After step 3 of the algorithm, the state of the first register will be in the Fourier basis, which is given by vectors
\begin{align}
    \ket{F_P(\omega)} = \frac{1}{\sqrt P}
    \sum_{j = 0}^{P - 1} e^{\im \omega j} \ket j,
\end{align}
where $\omega \in \Omega_P = \set{0, \frac{2\pi}{P}, \ldots, \frac{2\pi(P - 1)}{P}}$.
Then, the probability of obtaining $\tilde \theta$ as the estimation for $\theta$, when $\tilde\theta \neq \theta$,
is given by
\begin{align}
    \abs{\braket{F_P(\tilde{\theta}) | F_P(\theta)}}^2
    = \frac{\sin^2(P(\theta - \tilde{\theta}))}{
        P^2 \sin^2(\theta - \tilde{\theta})
    }.
    \label{eq:estimation-error}
\end{align}

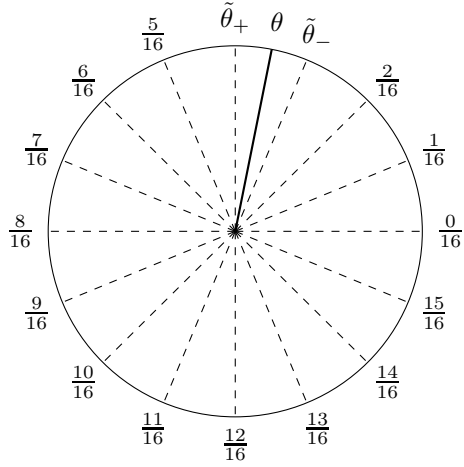
\begin{figure}[!htb]
    \centering
    \begin{tikzpicture}
        \def\scale{7}
        \def\P{16}
        \pgfmathsetmacro{\itemax}{\P - 1}
        \draw (0em, 0em) circle(\scale*1em);
        
        \foreach \p in {0, ..., \itemax}{
            \pgfmathsetmacro{\x}{cos(\p*2*pi/\P * 180/pi)}
            \pgfmathsetmacro{\y}{sin(\p*2*pi/\P * 180/pi)}
            \draw[dashed] (0em, 0em) -- (\x*\scale*1em, \y*\scale*1em);
            
            \pgfmathsetmacro{\thetam}{3}
            \pgfmathsetmacro{\thetambound}{\thetam - 1}
            \pgfmathsetmacro{\thetap}{\thetam+1}
            \pgfmathsetmacro{\thetapbound}{\thetap + 1}
            \ifthenelse{\p > \thetambound \AND \p < \thetapbound}{
                \ifthenelse{\p = \thetam}{
                    \def\var{$\tilde\theta_-$}
                }{
                \ifthenelse{\p = \thetap}{
                    \def\var{$\tilde\theta_+$}
                }{
                    \def\var{ERROR}
                }}
                
            }{
                \def\var{$\frac{\p}{\P}$}
            }
            \node at (\x*\scale*1.15em, \y*\scale*1.15em) {\var};
        }
        
        \pgfmathsetmacro{\x}{cos(3.5*2*pi/\P * 180/pi)}
        \pgfmathsetmacro{\y}{sin(3.5*2*pi/\P * 180/pi)}
        \draw [thick] (0em, 0em) -- (\x*\scale*1em, \y*\scale*1em);
        \node at (\x*\scale*1.15em, \y*\scale*1.15em) {$\theta$};
        
    \end{tikzpicture}
    \caption{Unitary circle subdivided into $P = 16$ parts.
             A good estimation for an angle $\theta$ is either
             $\tilde\theta_+$ or $\tilde\theta_-$,
             the angles that delimit the part in
             which $\theta$ is in.
             The $2\pi$ factor that multiplies all fractions was omitted.
             }
    \label{fig:phase-est}
\end{figure}

Intuitively, we can interpret the phase estimation algorithm as follows. The unitary circle is divided into $P$ equal parts, each part with angle $2\pi/P$. The estimation, $\tilde\theta$, can only assume a value in the set 
$\Omega_P$.
$\tilde\theta$ is a good estimate for $\theta$ if their difference is no greater than the part size, that is, $|\tilde\theta - \theta| \leq 2\pi/P$. In other words, if the value of $\theta$ is either rounded up to
$\tilde\theta_+ = \frac{2\pi}{P} \ceil{\frac{P \theta}{2\pi}}$
or rounded down to
$\tilde\theta_- =  \frac{2\pi}{P} \floor{\frac{P \theta}{2\pi}}$,
where
$\tilde\theta_+,\tilde\theta_-\in \Omega_P$.
It should be noted that $\tilde\theta_+$ and $\tilde\theta_- $ demarcate the part in which the angle $\theta$ resides -- see Figure~\ref{fig:phase-est}. The closer $\theta$ is to a value in
$\Omega_P$,
the higher the probability of obtaining a good estimate.
After using Eq.~\eqref{eq:estimation-error},
the probability of obtaining a good estimate is
\begin{equation}
    \abs{\braket{F_P(\tilde{\theta}_{+}) | F_P(\theta)}}^2+\abs{\braket{F_P(\tilde{\theta}_{-}) | F_P(\theta)}}^2\geq 8/\pi^2 .
\end{equation}
If $\theta$ is precisely in the middle of a part,
the probability of obtaining a good estimate is the smallest possible.
When $\theta \in \Omega_P$, we have $\tilde\theta = \theta$ with certainty.

\subsection{Quantum counting algorithm}
\label{sec:orig-count}

In a search problem, the goal is to find a specific element or solution within a given set of data. Consider a search space with $N$ elements and $k$ solutions (marked elements). 
Grover's search algorithm is defined by using a quantum oracle which can identify marked elements. Finding a marked element requires $O(\sqrt{N/k})$ calls to the oracle. However, the optimal number of steps of the algorithm depends on $k$ and this information is usually unknown. One way to find it is by using the quantum counting algorithm.  

The quantum counting algorithm is used to determine the number of solutions to a given problem. It is based on the quantum phase estimation algorithm. The algorithm starts with a superposition of all possible states in the second register and uses the Grover search operator as the controlled operator. The non-trivial eigenvalues of the Grover operator encode the number of marked elements. After running the quantum phase estimation algorithm, classical post-processing is performed to estimate these eigenphases, providing an estimation of the number of marked elements.

The Grover search operator is given by
\begin{align}
    G = (2\ketbra{d}{d} - I)R,    
\end{align}
where
\begin{align}
    \ket d = \frac{1}{\sqrt N}\sum_{j=0}^{N-1} \ket j.   
\end{align}
The quantum oracle $R$ can be defined as
\begin{align}
    R = I - 2 \sum_{j \in K} \ketbra j j,  
\end{align}
which flips the phase of states that belong to $K$ (set of marked elements).

Let $\ket{x_0}$ be the superposition of unmarked elements and $\ket{x_1}$ be the superposition of marked elements, that is, 
\begin{align*}
    \ket{x_0} &= \frac{1}{\sqrt{N - k}} \sum_{j \notin K}\ket j,\\
    \ket{x_1} &= \frac{1}{\sqrt k} \sum_{j \in K} \ket j,
\end{align*}
where $k = |K|$. In the two-dimensional plane spanned by $\ket{x_0}$ and $\ket{x_1}$,
$G$ acts as a rotation operator $R(\theta)$, whose expression is
\begin{align}\label{eq:Rtheta}
    R(\theta) = \matrx{
        \cos(\theta) & - \sin(\theta) \\
        \sin(\theta) & \cos(\theta)
    }.
\end{align}
The rotation angle $\theta$ is defined by
\begin{equation}
\cos\frac\theta 2 = \sqrt{\frac{N - k}{N}}\textrm{ and } \sin\frac \theta 2 =\sqrt{\frac k N}.
\label{eq:cos}
\end{equation}
The eigenvalues and eigenvectors of $R(\theta)$ are $e^{\pm \im\theta}$ and $\ket{\mp\im} = \pr{\ket 0 \mp \im\ket 1}/\sqrt 2$,
respectively. 
Therefore, the non-trivial eigenvectors of the Grover operator are $\ket{\mp\im_G} = (\ket{x_0} \mp\im \ket{x_1})/\sqrt 2$ with associated $e^{\pm\im \theta}$ eigenvalues. And we can express the uniform superposition $\ket{d}$ as
\begin{align}
    \ket d = \frac{e^{\im\theta/2}}{\sqrt 2}\ket{-\im_G}
             + \frac{e^{-\im\theta/2}}{\sqrt 2}\ket{+\im_G}.
\end{align}

By knowing the value of $\theta$, we can easily obtain the number of marked elements $k$. We simply do $k = N\sin^2\theta$, using Eq.~\eqref{eq:cos}. 
Passing $G$ and $\ket d$ as arguments to $\texttt{phase\_estimation}$ results in an estimation of $\pm \theta$ with an equal probability of $1/2$ for each. However, by definition, we know that $\theta \in [0, \pi]$. As a result, it becomes straightforward to discern whether $\theta$ or $-\theta$ was estimated, and subsequently make the necessary conversion. Due to its simplicity, the process of distinguishing between $\theta$ and $-\theta$ is often omitted.

The success probability of the quantum counting algorithm and the error analysis are given by Theorem 12 of~\cite{brassard2002quantum}, which states the following (after some adaptations to our notation).

\begin{theorem}
    Let $\tilde\theta$ be the estimation of the angle $\theta$ with $p$ digits of precision, obtained as the output of \textnormal{\texttt{phase\_estimation}}. Let $P = 2^p$.
    If $\theta \in \Omega_P$, then $\sin^2(\tilde\theta/2) = \sin^2(\theta/2)$ with certainty.
    Otherwise,
    \begin{align}
        \abs{ \sin^2 \frac{\tilde\theta}{2} - \sin^2\frac\theta 2 } \leq
        \frac{2\pi \sin\frac \theta 2 \cos\frac \theta 2}{P} + \frac{\pi^2}{P^2}
        \label{eq:teo-bhmt}
    \end{align}
    with probability at least $8/\pi^2$.
    In addition, the algorithm queries the oracle $P - 1$ times.
    \label{teo:bhmt}
\end{theorem}

\begin{corollary}
    Using the definition of $\cos\frac\theta 2$ and $\sin \frac\theta 2$ in Eq.~\eqref{eq:cos}, we obtain the error
    \begin{align}
    \left| \tilde k - k \right| \leq\frac{ 2\pi \sqrt{k(N - k)}}{P} + \frac{\pi^2 N}{P^2},
    \end{align}
    where $\tilde k$ is the estimation of the number of marked elements $k$.
    \label{cor:bhmt}
\end{corollary}

Let us take $P = O(\sqrt N)$. Then, if either $k \approx 1$ or $k \approx N$, the expected error is bounded by a constant. On the other hand, if $k \approx N/2$, the error is the largest possible, $|\tilde k - k| = O(\sqrt N)$.
This odd behavior is caused by the non-linearity of $\theta$ with respect to $k$, since $\theta = 2\arcsin\sqrt{k/N}$. Let $\theta^c = 2\arcsin\sqrt{(k+c)/N}$ for some appropriate positive integer constant $c$. The difference $|\theta^c - \theta|$ is smaller as $k$ approaches $N/2$ and larger as $k$ approaches $0$ or 
$N$. Consequently, considering $P = O(\sqrt N)$, there are more possible values for $\theta$ between $\theta_+$ and $\theta_-$ when $k \approx N/2$ than when $k \approx 0$ or $k \approx N$.
Therefore, the estimation of the quantum counting algorithm is better when the absolute difference between the number of marked and unmarked elements is larger. An error of $O(\sqrt N)$ when $k \approx N/2$ can also be obtained by classical randomized algorithms querying the oracle $\Omega(N)$ times~\cite{NC00}.

\section{Quantum walk on the complete bipartite graph}
\label{sec:evol-op}

Now we consider a coined quantum walk on a complete bipartite graph $\Gamma(V_0,V_1, E)$, with parts $V_0$ and $V_1$, and $E$ denoting the edge set~\cite{Wes00}. Let $|V_j| = N_j$, for $j \in \set{0, 1}$, and $N=N_0+N_1$.
A quantum walk on $\Gamma$ is defined by using operators acting on a Hilbert space ${\cal H}^{2|E|}$ spanned by
$\set{\ket{i}\ket{uv} : i \in \set{0, 1}, uv \in E}$.
We use two types of quantum registers: a single qubit and an $|E|$-dimensional qudit. This choice reflects the structure of the graph, where each edge $uv$ gives rise to two directed arcs. One arc goes from vertex $u$ to vertex $v$, and the other goes in the opposite direction, from vertex $v$ to vertex $u$. In this setup, the qubit's role is to indicate the direction of the arc, while the qudit represents the edge itself. In more specific terms, the state $\ket{0}\ket{uv}$ denotes the arc that goes from $V_0$ to $V_1$. In contrast, the state $\ket{1}\ket{uv}$ denotes the arc that goes from $V_1$ to $V_0$.
This approach ensures a one-to-one correspondence between our quantum representation and the structure of the graph.

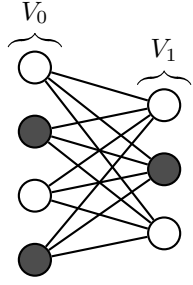
\begin{figure}[!htb]
\centering
\begin{tikzpicture}[thick,
  every node/.style={draw,circle},
  fsnode/.style={fill=white},
  ssnode/.style={fill=white},
  every fit/.style={ellipse,draw,inner sep=-2pt,text width=2cm},
  -,shorten >= 0pt,shorten <= 0pt
]

\begin{scope}[every node/.style={}]
\draw [decorate, 
    decoration = {calligraphic brace,
        raise=5pt,
        amplitude=5pt}] (-0.35,0.1) --  (0.35,0.1)
    node[pos=0.5,above=10pt,black]{$V_0$};
\draw [decorate, 
    decoration = {calligraphic brace,
        raise=5pt,
        amplitude=5pt}] (1.35,-0.4) --  (2.05,-0.4)
    node[pos=0.5,above=10pt,black]{$V_1$};
\end{scope}

\begin{scope}[start chain=going below,node distance=4mm]
\node[fsnode,on chain,minimum size=12pt,inner sep=0pt] (f1) {};
\node[fsnode,fill=black!70,on chain,minimum size=12pt,inner sep=0pt] (f2) {};
\node[fsnode,on chain,minimum size=12pt,inner sep=0pt] (f3) {};
\node[fsnode,fill=black!70,on chain,minimum size=12pt,inner sep=0pt] (f4) {};

\end{scope}

\begin{scope}[xshift=1.7cm,yshift=-0.5cm,start chain=going below,node distance=4mm]
\node[ssnode,on chain,minimum size=12pt,inner sep=0pt] (s1) {};
\node[ssnode,fill=black!70,on chain,minimum size=12pt,inner sep=0pt] (s2) {};
\node[ssnode,on chain,minimum size=12pt,inner sep=0pt] (s3) {};

\end{scope}

\draw (f1) -- (s1);
\draw (f1) -- (s2);
\draw (f1) -- (s3);
\draw (f2) -- (s1);
\draw (f2) -- (s2);
\draw (f2) -- (s3);
\draw (f3) -- (s1);
\draw (f3) -- (s2);
\draw (f3) -- (s3);
\draw (f4) -- (s1);
\draw (f4) -- (s2);
\draw (f4) -- (s3);
\end{tikzpicture}

\caption{A complete bipartite graph with $N_0=4$ and $N_1=3$. In this case, there are $k=3$ marked vertices, represented by the gray vertices. Two marked vertices in the set $V_0$ ($k_0 =2$) and one marked vertex in the set $V_1$ ($k_1 =1$).}
\label{fig:bip}
\end{figure}

Suppose we want to search for a marked vertex in the graph. Let $K_j$ be the set of marked vertices in $V_j$, $\bar K_j = V_j \setminus K_j$ be the set of unmarked vertices in $V_j$,
$K = K_0 \cup K_1$, $|K| = k$, and
$|K_j| = k_j$. An example of a complete graph with marked vertices is depicted in Fig.~\ref{fig:bip}.
The quantum walk search is performed by doing a query to an oracle followed by a step of the quantum walk. This process is described by the evolution operator $U = SCR$. The quantum oracle $R$ is given by
\begin{align}
    R = I - 2 \sum_{i = 0}^1 \sum_{u \in K_i} \sum_{v \sim u} \ket{i}\bra{i} \otimes \ket{uv}\bra{uv},
\end{align}
where $u\sim v$ denotes that $u$ and $v$ are adjacent. $R$ flips the sign of the amplitude of marked states. The product $SC$ performs a step of the standard quantum walk. The coin operator $C$, in this case, is the Grover coin
\begin{align}\label{eq:C}
    C = \bigoplus_{u}G_u,
\end{align}
where
\begin{align}\label{eq:Gv}
     G_u = 2\ketbra{d_u}{d_u} - I,
\end{align}
and
\begin{align}\label{eq:dv}
     \ket{d_u} = \frac{1}{\sqrt{\deg(u)}}\sum_{v\sim u}\ket i \ket{uv},
\end{align}
such that $u \in V_i$.
The shift operator is $S=X \otimes I$, and since $S^2=I$, it is a flip-flop shift operator.

The quantum walk dynamics has an invariant subspace spanned by the orthonormal vectors $\ket{K_0, K_1}$, $\ket{K_0, \bar K_1}$, $\ket{\bar K_0, K_1}$, $\ket{\bar K_0, \bar K_1}$,
$\ket{K_1, K_0}$, $\ket{K_1, \bar K_0}$, $\ket{\bar K_1, K_0}$, $\ket{\bar K_1, \bar K_0}$, where 
\begin{align}
    \ket{K_0, K_1} &= \frac{1}{\sqrt{\abs{K_0}\abs{K_1}}}
        \sum_{u \in K_0} \sum_{v \in K_1} \ket{0}\ket{uv}, \\
    \ket{K_1, K_0} &= \frac{1}{\sqrt{\abs{K_0}\abs{K_1}}}
        \sum_{u \in K_0} \sum_{v \in K_1} \ket{1}\ket{uv},
\end{align}
and the remaining ones are defined analogously. Note that $\ket{K_0, K_1}$ is a unit vector defined as the uniform superposition of the arcs that have their tails in $K_0$ and heads in $K_1$. The remaining vectors have analogous interpretations.

Using the definition of the Grover coin, Eq.~\eqref{eq:Gv}, and the flip-flop shift operator, we obtain
\begin{align}
    U \ket{K_0, K_1} = \pr{\frac{2k_1}{N_1} - 1} \ket{K_1, K_0} +
        \frac{2}{N_1} \sqrt{k_1 \pr{N_1 - k_1}} \ket{\bar K_1, K_0}.
\end{align}
Analogous expressions are obtained using the remaining vectors of the 8-dimensional basis. The action of $R$ in this subspace flips the sign of the states that have $K_0$ or $K_1$ in the first slot, otherwise, it makes no change. For example $R \ket{K_0, K_1} = - \ket{K_0, K_1}$ and
$R \ket{\bar K_0, K_1} = \ket{\bar K_0, K_1}$.
Therefore, we can express $U$ in the invariant basis as the 8-dimensional reduced operator
\begin{align}
    U_\text{RED} = \matrx{
        0 & \U(\theta_0) \\ \U(\theta_1) & 0
    },
\end{align}
where
\begin{align}
    \U ( x) = \matrx{
        \cos x  & -\sin x   &       0       &       0   \\
            0       &       0       & -\cos x    & \sin x \\ 
        -\sin x & -\cos x   &       0       &       0   \\
            0       &       0       & \sin x    & \cos x
    },
\end{align}
and the angles $\theta_j$ are defined such that
$\cos(\theta_j) = 1 - \frac{2k_j}{N_j}$ and $\sin(\theta_j) = \frac{2}{N_j} \sqrt{k_j(N_j - k_j)}$.

We consider the initial state of the quantum walk search as the uniform superposition of all arcs, that is, 
\begin{align}\label{eq:ket-d}
    \ket{d} &= \frac{1}{\sqrt{2|E|}} \sum_{i = 0}^1 \sum_{uv \in E} \ket{i}\ket{uv}.
\end{align}
Note that $\ket{d}$ belongs to the invariant subspace and can be written as

\begin{align}
     \ket{d} &= \frac{1}{\sqrt{2 N_0 N_1}} \left( \sqrt{\left|K_0\right| \left|K_1\right|}\ket{K_0, K_1}+ \sqrt{\left|K_0\right| \left|\bar K_1\right|}\ket{K_0, \bar K_1} + \cdots  \right) .
\end{align}

Previously, the coined quantum walk search on the complete bipartite graph was studied by Rhodes and Wong~\cite{rhodes2019quantum}. The eigenvalues and eigenvectors are obtained asymptotically for the general case. Following, we calculate its spectral decomposition.

\subsection{Eigenvalues and Eigenvectors}

The eigenvalues of $U_\text{RED}$ are the solutions $\lambda$ of 
\begin{align}
    \det(U_\text{RED} -\lambda I) = \det\pr{\lambda^2I - \U(\theta_0)\U(\theta_1)} = 0.
    \label{eq:evals}
\end{align}
Note that  $\U(\theta_0)\U(\theta_1) = R(\theta_0)^T \otimes R(\theta_1)$.
Using that $R(\theta)$ has eigenvalues $e^{\pm\im\theta}$
associated to eigenvectors $\ket{\mp\im}=(\ket{0}\mp\im\ket{1})/\sqrt{2}$, we conclude that $\U(\theta_0)\U(\theta_1)$ has four distinct eigenvalues $e^{\pm\im\pr{\theta_0 \pm \theta_1}}$.
Therefore, the eight distinct solutions of Eq.~\eqref{eq:evals} are $\lambda=e^{\pm\im\pr{\theta_0 \pm \theta_1}/2}$ and $\lambda=- e^{\pm\im\pr{\theta_0 \pm \theta_1}/2}$.

In order to obtain the eigenvectors,
suppose that $\ket\lambda = \ket{\lambda_0} \oplus \ket{\lambda_1}$
is a $\lambda$-eigenvector of $U_\text{RED}$.
Then,
\begin{align}
    U_\text{RED}\ket\lambda = \U(\theta_0)\ket{\lambda_1} \oplus \U(\theta_1)\ket{\lambda_0}
\end{align}
is equivalent to the system of equations
\begin{align}
    \begin{cases}
        \U(\theta_0)\ket{\lambda_1} = \lambda \ket{\lambda_0}, \\
        \U(\theta_1)\ket{\lambda_0} = \lambda \ket{\lambda_1}.
    \end{cases}
\end{align}
The first equation implies
$\ket{\lambda_0} = \frac 1 \lambda \U(\theta_0)\ket{\lambda_1}$.
Substituting into the second equation, we obtain
\begin{align}
    R(\theta_1)^T \otimes R(\theta_0) \ket{\lambda_1} = \lambda^2\ket{\lambda_1}.
\end{align}
That is, $\ket{\lambda_1}$ is a $\lambda^2$-eigenvector of $R(\theta_1)^T \otimes R(\theta_0)$.
Using the eigenvectors of $R(\theta)$, we obtain four eigenvectors $\ket{\lambda_1}$, which in compact form are 
$\ket{\lambda_1} = \ket{s_0 \im}\ket{s_1 \im}$ with eigenvalues
$e^{\im\pr{s_0\theta_1 - s_1\theta_0}}$ for $s_0=\pm 1$ and $s_1 =\pm 1$ independently.
Therefore, $\lambda = \pm e^{\im\pr{s_0\theta_1 - s_1\theta_0}/2}$.
After using the first equation to obtain $\ket{\lambda_0}$, we are able to list all eigenvectors $\ket{\lambda}$ of $U_\text{RED}$:
\begin{align}
    \ket{\mu_\pm} &= \frac{1}{\sqrt 8}\matrx{
        \pm e^{\im\sigma} & \mp\im e^{\im\sigma} & \pm\im e^{\im\sigma} & \pm e^{\im\sigma} &
        1 & -\im & \im & 1
    }^T,\label{eq:eigenvals-Uprime-1}\\    
    \ket{\mu_\pm^*} &= \frac{1}{\sqrt 8}\matrx{
        \pm e^{-\im\sigma} & \pm\im e^{-\im\sigma} & \mp\im e^{-\im\sigma} & \pm e^{-\im\sigma} &
        1 & \im & -\im & 1
    }^T,\\
    \ket{\sigma_\pm} &= \frac{1}{\sqrt 8} \matrx{
        \pm e^{\im\mu} & \pm\im e^{\im\mu} & \pm\im e^{\im\mu} & \mp e^{\im\mu}&
        1 & -\im & -\im & -1
    }^T,    \\
        \ket{\sigma_\pm^*} &= \frac{1}{\sqrt 8} \matrx{
        \pm e^{-\im\mu} & \mp\im e^{-\im\mu} & \mp\im e^{-\im\mu} & \mp e^{-\im\mu}&
        1 & \im & \im & -1
    }^T, \label{eq:eigenvals-Uprime-4} 
\end{align}
which are associated with the respective eigenvalues $\pm e^{\im\mu}$, $\pm e^{-\im\mu}$, $\pm e^{\im\sigma}$, and $\pm e^{-\im\sigma}$, where
$\mu = (\theta_0 + \theta_1)/2$ and $\sigma = (\theta_0 - \theta_1)/2$.
These eigenvectors form an orthonormal basis.

\section{Counting on the Complete Bipartite Graph}\label{sec:counting-complete-bipartite}

As described in Section~\ref{sec:orig-count}, the original quantum counting algorithm estimates an eigenphase of the evolution operator of Grover's algorithm, which results in the number of marked elements. In this section, we describe a similar approach for calculating the number of marked vertices in a graph using the quantum phase estimation algorithm with an evolution operator of a quantum walk that encodes the number of marked vertices in its eigenvalues.

Our objective is to ascertain the total number of marked vertices, denoted as $k_0 + k_1$, in the complete bipartite graph. Recall the spectral decomposition of the quantum walk search operator we discussed in the preceding section. It's important to note that determining $\theta_j$ allows us to also calculate $k_j$. This relationship is expressed by the equation
\begin{align}
    \cos(\theta_j) = 1 - \frac{2k_j}{N_j}
    &\implies k_j = N_j \sin^2 \frac{\theta_j}{2} .
\end{align}
Given this, the evolution operator of the quantum walk search, represented as $U = SCR$, becomes a prime candidate for our analysis.

First, we calculate the probability of estimating each eigenphase when
$U$ and $\ket{d}$, given by Eq.~\eqref{eq:ket-d}, are inputs for the phase estimation algorithm.
In the eigenbasis of $U$, we have
\begin{align}
\ket{d} = \sum_\lambda \braket{\lambda | d} \ket\lambda,
\end{align}
where $\ket\lambda$ represents the eigenvectors described in Eqs~\eqref{eq:eigenvals-Uprime-1} to~\eqref{eq:eigenvals-Uprime-4}. The modulus square of the coefficients $\braket{\lambda | d}$ are 
\begin{align}
\abs{\braket{\mu_\pm | d}}^2 &= \frac{1}{8} \paren{1 \pm \cos(\sigma)},\\
\abs{\braket{\sigma_\pm | d}}^2 &= \frac{1}{8} \paren{1 \pm \cos(\mu)}.
\end{align}
Using the trigonometric identities $1 + \cos x = 2\cos^2(x / 2)$ and $1 - \cos x = 2 \sin^2(x/2)$, we obtain the probability of estimating an eigenphase. The results are compiled in Table~\ref{tbl:estim-prob}.
Note that there is an equal probability of estimating the positive and negative phases.

\begin{table}[htb]
    \centering
    \def\arraystretch{1.5}
    \begin{tabular}{|c|c|}
        \hline
        phase & probability \\ \hline
        $\pm\mu$ & $\frac{1}{4}\cos^2\paren{\sigma / 2}$
        \\ \hline
        $\pm(\pi - \mu)$ & $\frac{1}{4}\sin^2\paren{\sigma / 2}$
        \\ \hline
        $\pm\sigma$ & $\frac{1}{4}\cos^2\paren{\mu / 2}$
        \\ \hline
         $\pm(\pi - \sigma)$ & $\frac{1}{4}\sin^2\paren{\mu / 2}$
        \\ \hline
    \end{tabular}
    \caption{Probability of estimating the eigenphases of $U$ when the initial state is the uniform superposition of all arcs of the graph.}
    \label{tbl:estim-prob}
\end{table}

Running the phase estimation algorithm once
yields an estimation of one out of eight possible eigenphases.
We could estimate $\mu$ and $\sigma$ in order to calculate $\theta_0$ and $\theta_1$.
However, multiple executions of the phase estimation algorithm would be required to correctly identify $\mu$ and $\sigma$.
On top of that, it is not a trivial task to distinguish between the eight eigenphases.
To avoid the distinguishing task, we will tackle the counting problem in cases where marked vertices exist in only one part of the bipartite graph.
Then, we extend this solution to handle the general case.

\subsection{Marked vertices in one part}

Consider the case where all the marked vertices are located in one part of the bipartite graph.
For instance, suppose the marked vertices are located in $V_0$ ($K \subseteq V_0$).
In this case, $k_1 =0$ and $\theta_1 = 0$, which implies $\mu = \sigma = \theta_0 / 2$. On the other hand, when we consider the second possibility, that is $K \subseteq V_1$, we have $k_0 =0$ and $\theta_0 = 0$, which implies $\mu = \theta_1 / 2$ and $\sigma=-\theta_1/2$.
Hence, in both cases, $U$ has only four eigenphases:
$\pm\theta_j / 2$, and $\pm(\pi - \theta_j/2)$.
Additionally, since $0 \leq \theta_j/2 \leq \pi/2$,
we can convert any other eigenphase to $\theta_j / 2$ by
performing a conditional reflection on the real axis
($x \to -x$ if $\pi < x < 2\pi$),
followed by a conditional reflection on the imaginary axis
($x \to \pi - x$ if $x > \pi/2$).
For simplicity, we will omit these reflections.

We proceed by executing the phase estimation algorithm, using the operator $U$ and the state $\ket{d}$ as inputs. This algorithm is designed to yield a close approximation, denoted as $\tilde{\theta}_j / 2$, with a probability of at least $8/\pi^2$. However, to accurately estimate $k_j$, it's essential to identify the specific part $j$ where the marked vertices reside. This identification is necessary to accurately estimate $k_j$ by multiplying with the correct part size $N_j$. We then estimate $k_j$ using the formula $\tilde k_j = N_j \sin^2(\tilde\theta_j/2)$.
The details of this procedure are depicted in Algorithm~\ref{alg:count-in-part}.

\begin{algorithm}[!htb]
    \caption{\texttt{partial\_count($p, R, j, N_0, N_1$)}}
    \label{alg:count-in-part}   
    \KwIn{
        The number of qubits in the first register $p$,
        the part $j$ where the marked vertices are located,
        and the graph $K_{N_0, N_1}$.}
    \KwOut{$\tilde{k}_j$. }
    Prepare the uniform state $\ket d$ and the evolution operator $U = SCR$\;
    $\tilde\theta_j / 2 \gets $ \texttt{phase\_estimation($p, U, \ket d$)}\;
    $\tilde k_j \gets N_j \sin^2(\tilde\theta_j / 2)$\;
    \Return $\tilde k_j$.
\end{algorithm}

To determine the error associated with the count of marked vertices, we cannot directly utilize Corollary \ref{cor:bhmt}. The reason for this is that the definitions of $\cos(\theta_j/2)$ and $\sin(\theta_j/2)$ in the context of the complete bipartite graph differ from those in Grover's algorithm. However, we can apply Theorem~\ref{teo:bhmt}, which leads us to establish Theorem~\ref{teo:bipartite}.

\begin{theorem}
        \label{teo:bipartite}
    Consider a bipartite graph $K_{N_0, N_1}$ where all marked vertices are located in one part $j$. By using algorithm \textnormal{\texttt{partial\_count}}, the error of the estimation $\tilde k_j$ of the number of marked vertices $k_j$ in
    $V_j$ is upper bounded by
    \begin{align}
        \abs{ \tilde k_j - k_j } \leq \frac{2\pi \sqrt{k_j(N_j - k_j)}}{P} +
        \frac{\pi^2 N_j}{P^2}
    \end{align}
    with success probability at least $8/\pi^2$. If 
    $\theta_j \in \Omega_P$, then $\tilde k_j = k_j$ with certainty. Moreover, the algorithm does $P-1$ queries to the oracle.
    
    \begin{proof}
        By multiplying Eq.~\eqref{eq:teo-bhmt} by $N_j$,
        using the trigonometric identity
        $\sin\frac x 2 \cos \frac x 2 = \frac{\sin x}{2}$,
        and the definition of $\sin(\theta_j)$,
        we obtain the given upper bound.
        The success probability and number of queries to the oracle come directly from Theorem \ref{teo:bhmt}.
    \end{proof}
\end{theorem}

Interestingly, Theorem~\ref{teo:bipartite} yields a result that mirrors that of Corollary~\ref{cor:bhmt}. Consequently, this allows us to count the number of marked vertices in a single part of the graph, adhering to the same error bound established by the original quantum counting algorithm. We leverage this result to the more general scenario, where marked vertices may reside in both parts of the graph.

\subsection{General case}

In the preceding subsection, we illustrated the method for counting marked vertices within a part of the bipartite graph. However, in the general scenario, marked vertices may be distributed across both parts. To address this case, we modify the oracle to exclusively mark vertices in one part at a time. This adaptation effectively reduces the general problem to the subproblem we tackled earlier. By employing this strategy, we can resolve the counting problem efficiently by executing the phase estimation algorithm twice

We construct two distinct oracles based on $R$, as follows:
\begin{align}
    R_j \ket{i}\ket{uv} = \begin{cases}
        R \ket i\ket{uv} \ &\textnormal{if } i = j, \\
        \ket i\ket{uv} &\textnormal{otherwise,}
    \end{cases}
\end{align}
where $j \in \set{0, 1}$. In the next subsection, we delve into the specifics of implementing the modified oracles. $R_j$ is designed to specifically flip the phase of vertices in $K_j$. Following this, we will outline the implementation of the original oracle. Then, we'll detail the process for implementing $R_j$.

\subsubsection{Implementation of the modified oracles}

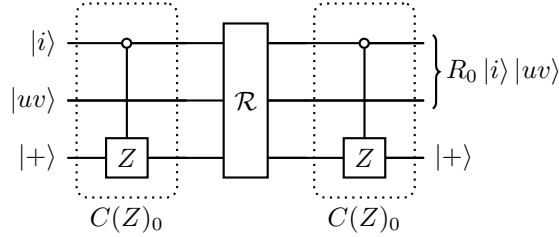
\begin{figure}[htb]
    \centering
    \begin{quantikz}
        \lstick{$\ket{i}$}
            & \octrl{2} \gategroup[3,style={dotted, rounded corners, inner xsep=0.75em},
                label style={label position=below, anchor=north, yshift=-0.5em}
              ]{$C(Z)_0$}
            & \qw
            & \gate[3]{\mathcal{R}}
            & \qw
            & \octrl{2} \gategroup[3,style={dotted, rounded corners, inner xsep=0.75em},
                label style={label position=below, anchor=north, yshift=-0.5em}
              ]{$C(Z)_0$}
            & \rstick[2]{$R_0\ket{i}\ket{uv}$}\qw
        \\
        \lstick{$\ket{uv}$}
            & \qw & \qw & & \qw & \qw & \qw
        \\
        \lstick{$\ket{+}$}
            & \gate{Z} & \qw & & \qw & \gate{Z} & \rstick{$\ket{+}$} \qw
    \end{quantikz}
    \caption{
        Circuit that implements $\mathcal{R}_0$, where
        $\ket{i}\ket{uv}$ is a state of the computational basis, and $\ket{+}$ is the initial state of an ancilla qubit.
    }
    \label{fig:modified-oracle}
\end{figure}

To implement $R$, an ancilla qubit $\ket\varphi$ is required. When the ancilla qubit is set to $\ket\varphi=\ket -$,
$R$ inverts the phase of the marked vertices. Let $\mathcal{R}$ and $\mathcal{R}_j$ represent extensions of $R$ and $R_j$ that act on the expanded Hilbert space. We define $\mathcal{R}$ as
\begin{align}
    \mathcal{R} \ket{i}\ket{uv} \ket\varphi = \begin{cases}
        \ket i \ket{uv} X \ket \varphi &\textnormal{if } u \textnormal{ or } v \in K_i,\\
        \ket i\ket{uv}\ket{\varphi} &\textnormal{otherwise.}
    \end{cases}
\end{align}
Next, we define two controlled $Z$ gates with the target on the ancilla qubit, which activate when the walker is in $V_j$,
\begin{align}
    C(Z)_j\ket i\ket{uv}\ket\varphi = \begin{cases}
        \ket i\ket{uv} Z\ket{\varphi} &\textnormal{if } i = j,\\
        \ket i\ket{uv}\ket{\varphi} &\textnormal{otherwise} .
    \end{cases}
\end{align}
Here, $C(Z)_0$ is a controlled $Z$ gate activated by $\ket 0$ and $C(Z)_1$ is activated by $\ket 1$.
$\mathcal{R}_j$ is implemented using
\begin{align}
    \mathcal{R}_j = C(Z)_j \mathcal{R}\, C(Z)_j,
\end{align}
with the ancilla qubit initially set to $\ket +$.
The circuit for $\mathcal{R}_0$ is illustrated in Fig.~\ref{fig:modified-oracle}.

In the implementation of $\mathcal{R}_j$,
$C(Z)_j$ maps $\ket +$ to $\ket -$
(and vice-versa) when the walker is in $V_j$,
$\mathcal{R}$ flips the phase only for the vertices in $K_j$.
On the other hand,
when the walker is not in $V_j$,
$C(Z)_j$ is not activated
and $\mathcal{R}$ does not flip the phase.

\subsubsection{Counting algorithm}

To solve the counting problem, we use the reduced versions $R$ and $R_j$ in Algorithm~\ref{alg:count-in-part}. For a practical implementation on a quantum computer, the expanded versions $\mathcal{R}$ and $\mathcal{R}_j$ would be necessary. This algorithm must be executed twice: first with $R_0$ and then with $R_1$. Subsequently, we calculate the estimated total number of marked elements, expressed as $\tilde k = \tilde k_0 + \tilde k_1$. The entire process is detailed in Algorithm~\ref{alg:count-bipartite}.

\begin{algorithm}[!htb]
    \caption{\texttt{full\_count($p, R, N_0, N_1$)}}
    \label{alg:count-bipartite}   
    \KwIn{
        The number of qubits in the first register $p$,
        the oracle $R$ and the graph $K_{N_0, N_1}$.}
    \KwOut{$\tilde{k}$. }
    Prepare the modified oracles $R_j = C(Z)_j R \, C(Z)_j$ for $j \in \set{0, 1}$\;
    $\tilde k_0 \gets$ \texttt{partial\_count($p, R_0, 0, N_0, N_1$)}\;
    $\tilde k_1 \gets$ \texttt{partial\_count($p, R_1, 1, N_0, N_1$)}\;
    \Return $\tilde k_0 + \tilde k_1$.
\end{algorithm}

\begin{theorem}
        \label{teo:bipartite-general}
    Consider a bipartite graph $K_{N_0, N_1}$. By using algorithm \textnormal{\texttt{full\_count}}, the error of the estimation $\tilde k$ of the number of marked vertices $k$ is upper bounded by
    \begin{align}
        \abs{\tilde k - k} \leq
    \frac{2\pi}{P} \pr{
       \sqrt{k_0(N_0 - k_0)} +
        \sqrt{k_1(N_1 - k_1)}
    }
    + \frac{\pi^2 N}{P^2}
    \end{align}
    with a success probability of at least $0.65$. Moreover, the algorithm does $2P-2$ queries to the oracle.
    
    \begin{proof}
        The results are obtained by applying Theorem~\ref{teo:bipartite} twice.
    \end{proof}
\end{theorem}

Similar to the original counting algorithm, by setting $P = O(\sqrt{N})$, we obtain an error upper bound of $O(\sqrt{N})$. This is accomplished by making $O(\sqrt{N})$ calls to the oracle.

\section{Final Remarks}\label{conclusion}

In this work, we have extended the quantum counting algorithm by employing the quantum walk search evolution operator on a bipartite graph. In this way, we are able to count the number of marked vertices in a complete bipartite graph. By calculating the spectral decomposition of the quantum walk search evolution operator on the complete bipartite graph, we obtained that its eigenvalues encode the number of marked vertices in the graph. The idea is to use it as input to the quantum phase estimation algorithm.

When the marked vertices are located in one part of the graph, a straightforward use of the phase estimation algorithm allows us to determine the number of marked vertices with the same error bounds as the original quantum counting algorithm.  In the general scenario, where marked vertices may reside in both parts of the graph, we modify the oracle by adding a few controlled $Z$ gates. This modification allows us to independently estimate the number of marked vertices in each part. Such an approach is crucial for distinguishing the eigenphases that result from applying the phase estimation procedure, which might otherwise be indistinguishable. In this more complex case, the error bounds remain comparable to those of the original counting algorithm.

This method can be extended to other types of graphs as well. The key requirement is that the evolution operator must encode the number of marked elements in its eigenphases, which then can be used as input for the quantum phase estimation algorithm. The challenging aspect lies in accurately distinguishing the output eigenphases to correctly estimate the number of marked vertices. Furthermore, it raises an interesting question: Does this counting algorithm maintain error bounds that are comparable to those of the original counting algorithm?

\section*{Acknowledgments}

The work of G. A. Bezerra was supported by CNPq grant n.~146193/2021-0. The work of R. A. M. Santos was supported by ERDF project n.~1.1.1.2/VIAA/1/16/002 and Latvian Quantum Initiative under European Union Recovery and Resilience Facility project no. 2.3.1.1.i.0/1/22/I/CFLA/001. The work of R. Portugal was supported by FAPERJ grant number CNE E-26/200.954/2022, and CNPq grant numbers 308923/2019-7 and 409552/2022-4. The authors have no competing interests to declare that are relevant to the content of this article. All data generated or analyzed during this study are included in this published article.

\bibliographystyle{alpha}
\bibliography{bib}

\end{document}